\documentclass[sigplan,nonacm]{acmart}

\usepackage{graphicx}
\usepackage{url}
\usepackage[flushleft]{threeparttable}
\newcommand{\ignore}[1]{}

\usepackage[caption=false]{subfig}
\usepackage{stfloats}
\usepackage{enumitem}
\usepackage{stfloats}
\usepackage{multirow}
\usepackage{hhline}
\usepackage{array}
\usepackage{tabularx}
\usepackage{xcolor}
\usepackage{threeparttable}

\usepackage{graphicx}
\usepackage{textcomp}

\definecolor{aliceblue}{rgb}{0.94, 0.97, 1.0}
\usepackage{xcolor}
\usepackage{tcolorbox}
\tcbset{width=1\columnwidth,boxrule=1pt,colback=aliceblue,arc=1pt,left=2pt,right=2pt,boxsep=0.5pt}

\usepackage[flushleft]{threeparttable}
\usepackage{enumitem}

\AtBeginDocument{%
  }

\setcopyright{acmlicensed}

\usepackage{hyperref}
\begin{document}

\title[]{Enigma: Application-Layer Privacy for \\Quantum Optimization on Untrusted Computers}

\author{Ramin Ayanzadeh}
\affiliation{%
  \institution{University  of Colorado Boulder}
  \city{Boulder}
  \state{CO}
  \country{USA}
}

\author{Ahmad Mousavi}
\affiliation{%
  \institution{American University}
  \city{Washington}
  \state{DC}
  \country{USA}
}

\author{Amirhossein Basareh}
\affiliation{%
  \institution{Purdue University}
  \city{West Lafayette}
  \state{IN}
  \country{USA}
  }

\author{Narges Alavisamani}
\affiliation{%
  \institution{Georgia Tech}
  \city{Atlanta}
  \state{GA}
  \country{USA}
}

\author{Kazem Taram}
\affiliation{%
  \institution{Purdue University}
  \city{West Lafayette}
  \state{IN}
  \country{USA}
}


\renewcommand{\shortauthors}{}

\begin{abstract}

The Early Fault-Tolerant (EFT) era is emerging, where modest Quantum Error Correction (QEC) can enable quantum utility before full-scale fault tolerance.  
Quantum optimization is a leading candidate for early applications, but protecting these workloads is critical since they will run on expensive cloud services where providers could learn sensitive problem details.
Experience with classical computing systems has shown that treating security as an afterthought can lead to significant vulnerabilities. 
Thus, we must address the security implications of quantum computing before widespread adoption. 
However, current Secure Quantum Computing (SQC) approaches, although theoretically promising, are impractical in the EFT era: blind quantum computing requires large-scale quantum networks, and quantum homomorphic encryption depends on full QEC—neither of which is expected in the near term.

We propose {\em application-specific SQC}, a principle that applies obfuscation at the application layer to enable practical deployment while remaining agnostic to algorithms, computing models, and hardware architectures.  
We present {\em Enigma}, the first realization of this principle for quantum optimization.  
Enigma integrates three complementary obfuscations: \emph{ValueGuard} scrambles coefficients, \emph{StructureCamouflage} inserts decoys, and \emph{TopologyTrimmer} prunes variables.  
These techniques guarantee recovery of original solutions, and their stochastic nature resists repository-matching attacks.  
Evaluated against seven state-of-the-art AI models across five representative graph families, even combined adversaries, under a conservatively strong attacker model, identify the correct problem within their top five guesses in only 4.4\% of cases.  
The protections come at the cost of problem size and $T$-gate counts increasing by averages of 1.07× and 1.13×, respectively, with both obfuscation and decoding completing within seconds for large-scale problems.

\end{abstract}



\begin{teaserfigure}
  \vspace{0.3in}
  \end{teaserfigure}


\maketitle

\section{Introduction}

\begin{figure*}[t]
	\centering
	\includegraphics[width=0.85\textwidth]{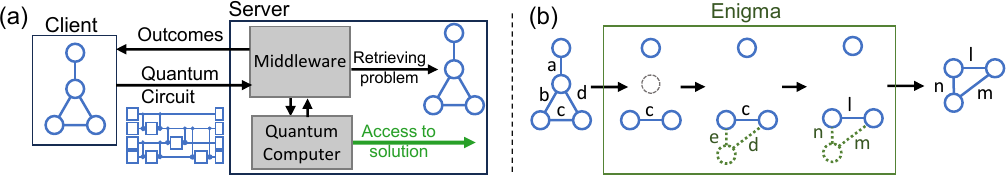}
    \caption{		
		(a)~Untrusted servers can infer the original optimization problem from the circuit.  
		(b)~Enigma workflow. After identifying the input graph type, Enigma applies TopologyTrimmer to prune nodes, StructureCamouflage to insert decoy nodes and edges in a graph-aware manner, ValueGuard to scramble coefficients, and StructureSanitizer to remove residual footprints.  
	}
	\label{fig:intro_fig}
\end{figure*}

Quantum optimization has the potential to significantly enhance critical sectors by addressing problems beyond classical limits. 
For instance, it can accelerate drug discovery and protein design in healthcare~\cite{emani2021quantum,robert2021resource,boulebnane2023peptide,cordier2021biology,bulancea2019quantum,sarkar2021quaser}, 
identify novel materials for energy storage and carbon capture~\cite{greene2022modelling,montgomery2025multi,paudel2022quantum,bauer2020quantum}, 
optimize logistics for greener supply chains~\cite{azad2022solving,bentley2022quantum,fitzek2024applying,bentley2022quantum,dalyac2021qualifying}, 
and improve power grid resilience and stability~\cite{colucci2023power,blenninger2024quantum,saevarsson2022quantum}. 
As we enter \emph{Early Fault-Tolerant (EFT)} era~\cite{preskill2025beyond,katabarwa2024early,kiss2025early,dangwal2025variational,he2024parameter}, modest levels of quantum error correction (QEC)~\cite{calderbank1996good, dennis2002topological, fowler2012surface, kitaev1997quantum, landahl2011fault, shor1995scheme, steane1996multiple} can support practical computations but full-scale fault tolerance remains out of reach. This make identifying applications that fit EFT crucial. 
Quantum optimization is emerging as one of the leading candidates, offering opportunities for near-term quantum utility under realistic EFT constraints~\cite{he2024parameter,preskill2025beyond}. 
Thus, the security and privacy of these valuable workloads become critical. 

Optimization workloads often encode highly sensitive intellectual property~\cite{cordier2021biology,bulancea2019quantum,sarkar2021quaser, emani2021quantum,robert2021resource,baker2022wasserstein,Barkoutsos_2020,egger2020quantum}, making their protection essential for practical quantum utility.  
As shown in Fig.~\ref{fig:intro_fig}(a), untrusted cloud servers can access and misuse such information, exposing advanced technologies and strategies across sectors.  
In pharmaceutical discovery, for example, the optimization formulation of a protein design problem directly reveals the therapeutic target under investigation, risking billions of dollars of R\&D investment if compromised.  
In energy and logistics, leaked models can disclose proprietary infrastructure or strategic plans.  
Finance provides another critical example: if an adversary reconstructs a portfolio optimization instance, they can mount \emph{benign imitation attacks} by free-riding on the firm’s strategy, or \emph{malicious manipulation attacks} by trading against revealed positions, such as shorting assets the firm holds long or buying those it has shorted.  
Even partial leakage of problem structure or coefficients can enable front-running and price manipulation, highlighting that optimization workloads embody strategic assets rather than personal identifiers.  

Lessons from classical computing underline the risks of overlooking security in early design stages, which can bring long-lasting and expensive-to-fix issues. 
For quantum computing, it is never too early to consider security implications, especially as sensitive workloads will increasingly depend on untrusted service providers.

Unfortunately, existing \emph{Secure Quantum Computing (SQC)} schemes are not deployable in the EFT era.  
Blind quantum computing requires quantum networks, and quantum homomorphic encryption assumes full QEC, neither of which will be available in the near term~\cite{fitzsimons2017private,broadbent2009universal,barz2012demonstration, li2021blind}.  
By contrast, the idea of injecting decoy operations on the client side and attenuating them within a trusted control environment is reasonable in the Noisy Intermediate-Scale Quantum~\cite{preskillNISQ} era with bare qubits and gates, but  becomes challenging as we move to EFT with logical operations under QEC.  

Meanwhile, the quantum ecosystem is evolving rapidly, with no clear technological or algorithmic winner.  
Computing models have diversified from gate-based and adiabatic~\cite{albash2018adiabatic} to measurement-based~\cite{briegel2009measurement} and, more recently, fusion-based paradigms~\cite{bartolucci2023fusion}.  
On the algorithmic side, while the Quantum Approximate Optimization Algorithm (QAOA)~\cite{farhi2014quantum} and the Variational Quantum Eigensolver (VQE)~\cite{tilly2022variational, kandala2017hardware} are leading candidates for demonstrating quantum utility, new algorithms may soon emerge.  
In such a volatile landscape, tying SQC to a specific technology, model, or algorithm risks limited transferability, motivating approaches that remain adaptable as the stack continues to evolve.

The near-term impracticality of prior work arises from its pursuit of a one-size-fits-all approach to securing quantum programs, which forces reliance on emerging technologies or incurs exponential overhead.  
To address this, we propose \textbf{application-specific SQC}, a design principle that brings obfuscation directly to the application layer, before a quantum program is generated.  
By aligning protections with workload structure, this approach enables practical deployment on current and near-term systems while remaining agnostic to algorithm, computing model, qubit technology, and hardware backend as the quantum stack evolves.

We present \emph{Enigma}, the first instantiation of application-specific SQC, realized in the context of quantum optimization.  
Quantum optimization workloads expose four critical assets: problem coefficients, variables, graph structure, and solution.  
To protect these assets, we introduce three obfuscation schemes:  
(1)~\emph{ValueGuard}, which perturbs coefficients to induce random bit-flips in outcomes, safeguarding both coefficients and solution;  
(2)~\emph{StructureCamouflage}, which adds decoy nodes and edges indistinguishable from real ones to conceal variables and topology; and  
(3)~\emph{TopologyTrimmer}, which removes a subset of primary nodes and their incident edges to obscure graph structure.  

All three obfuscations preserve the optimization form: both input and output remain problems of the same kind.  
This uniformity allows each scheme to map one instance into another without altering the interface.  
As a result, they can be applied in any order, reused multiple times, and extended with future designs that follow the same principle.  
For every obfuscation, we guarantee that once the server returns the solution of the obfuscated problem, the client can recover the solution of the original problem.  
Decoding is trivial for the client, but nontrivial for the server or any adversary without access to the decrypting information.  

Enigma applies these obfuscations strategically to protect workloads in a topology-aware manner.  
Before applying transformations, it identifies the input graph type and adjusts parameters to tailor the protection strategy.  
For instance, in a fully connected graph (SK type), decoy nodes must connect to all primary and decoy nodes to preserve full connectivity; otherwise, an adversary can isolate the original problem by extracting the largest clique.  
Similarly, in power-law graphs, randomly inserting decoy nodes may expose high-degree hubs, revealing the original structure.  

Figure~\ref{fig:intro_fig}(b) illustrates Enigma’s workflow.  
After identifying the graph type, Enigma first applies TopologyTrimmer to prune up to ten nodes and their incident edges.  
It then applies StructureCamouflage to insert decoy nodes and edges in a graph-type–aware manner.  
Next, ValueGuard scrambles primary and decoy coefficients to further obscure the problem.  
Finally, StructureSanitizer removes residual footprints, such as isolated nodes that could otherwise reveal the original graph.  
Because Enigma incorporates stochasticity, each run produces a distinct obfuscated output, making it nontrivial for an adversary to match instances against a repository of known problems.

To evaluate Enigma, we developed a new methodology, since no established metric exists for SQC in the systems community and classical security measures do not apply.  
We empirically evaluate the effectiveness of Enigma's obfuscation against powerful adversaries.
To that end, we challenge Enigma against seven state-of-the-art ML models trained on a large repository of graphs and their obfuscations spanning five representative families: 3-regular, fully connected, two power-law variants, and random graphs.  
Each model output a ranked list of five candidates as the potential original graphs, and we deemed an attack successful only if the actual original graph appeared among them.  
Across these experiments, Enigma consistently provided strong protection: even when combining all eight models, adversaries could recover the original problem in only 4.4\% of cases.

Enigma introduces three main sources of overhead. 
First, obfuscation inflates the problem size: the number of qubits grows by an average of 1.07x (and up to 1.16x), reflecting the cost of decoys. 
Second, this expansion translates into higher QEC resource demands, with $T$-gate counts increasing by an average of 1.13x (and up to 1.4x). 
Finally, the runtime overhead of Enigma is modest: both obfuscation and decoding complete in seconds even for large-scale inputs, making preprocessing negligible compared to quantum execution.

\vspace{0.5cm}
\noindent
Overall, this paper makes the following contributions:
\begin{itemize}[leftmargin=*] 

	\item Introduce \textbf{application-specific secure quantum computing}, a principle that applies obfuscation at the application layer for practical deployment on near-term systems.  
	\item Present \textbf{Enigma}, the first application-specific SQC framework for quantum optimization.  
\item Devised \emph{ValueGuard}, \emph{StructureCamouflage}, and \emph{TopologyTrimmer} to protect coefficients, variables, and graph structure while ensuring solution recoverability.  
\item Propose an evaluation methodology that leverages adversarial AI/ML models to assess obfuscation efficacy.  

\end{itemize}

\begin{figure}[h]    
    \centering        
	\includegraphics[width=0.9\columnwidth]{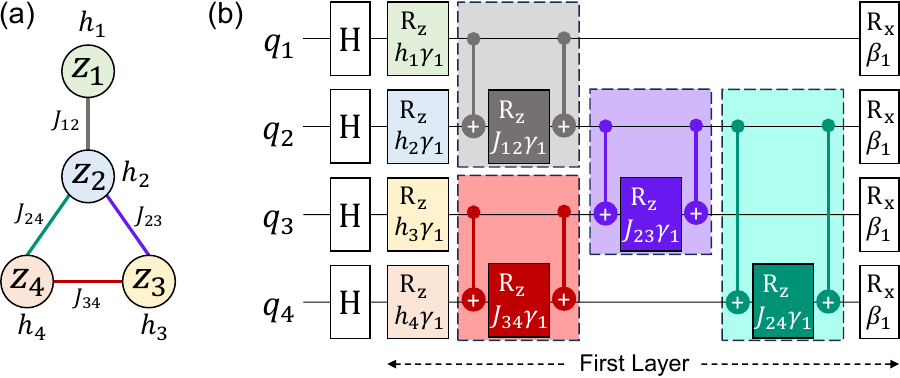}
    \caption{
    QAOA example:    
    (a)~Graph representation of the Ising model; 
        (b)~coresponding QAOA circuit with one layer.
        Every edge in the problem graph is represented by two two-qubit gates wrapping a single-qubit $R_z$ rotation, a pattern that an adversary can exploit to recover the original problem from the given circuit.
        }    
    \label{fig:qaoa_example_background}
\end{figure}

\section{Background and Motivation}

\subsection{Quantum Optimization}

Preskill’s Noisy Intermediate-Scale Quantum (NISQ) era~\cite{preskillNISQ} explored near-term devices for applications such as optimization, machine learning, and simulation through error mitigation and noise resilience~\cite{preskillNISQ,huang2021information,farhi2014quantum}.  
With rapid hardware advances, we are entering the \emph{Early Fault-Tolerant} (EFT) era, where \emph{MegaQuOp} systems with $10^4$--$10^5$ qubits can execute millions of reliable operations using Quantum Error Correction (QEC) with shorter code distances for deeper and more stable circuits~\cite{preskill2025beyond}.  
A central challenge is identifying algorithms whose resource demands match EFT constraints, and quantum optimization, particularly Ising-model solvers, offers strong potential in this regime.

\subsubsection{The Ising Model}
The Ising model, also referred to as the Ising Hamiltonian, is one of the most influential and enduring mathematical formulations in physics and computation.  
It has served as a foundational tool for understanding collective behavior in complex systems for over a century.  
The model describes a set of binary spin variables $\mathbf{z}_i \in \{-1, +1\}$ that interact through linear biases $\mathbf{h}_i \in \mathbb{R}$ and pairwise couplings $J_{ij} \in \mathbb{R}$.  
The objective is to approximate the ground state, that is, the spin configuration that minimizes the total energy:  

\begin{equation}
	f(\mathbf{z}) =
    \sum_{i=1}^{n} \mathbf{h}_i \mathbf{z}_i +
		\sum_{i=1}^{n} \sum_{j=i+1}^{n} J_{ij} \mathbf{z}_i \mathbf{z}_j.
	\label{eqn:Ising}
\end{equation}

In classical computing, many NP problems, such as graph coloring and the traveling salesman problem, are inter-reducible, with Boolean satisfiability (SAT) serving as a unifying abstraction.  
The Ising model may play a similar role in quantum computing.  
For most problems, designing efficient quantum algorithms and corresponding circuits is nontrivial.  
In contrast, the Ising model maps naturally onto hardware such as Quantum Annealers (QAs) and can be directly and systematically expressed as parameterized quantum circuits.  
This compatibility can position the Ising model as a \emph{Rosetta Stone} for translating diverse computational problems into quantum form, making advances in approximating its ground state immediately beneficial across applications.

\subsubsection{QAOA: A Prime Quantum Ising Solver}
QAOA approximates the ground state of an Ising model using a parametric quantum circuit with repeated layers.  
Inside each layer, for each edge in the problem graph [Fig.~\ref{fig:qaoa_example_background}(a)], the circuit applies two $CX$ gates and an $R_z$ rotation, and ends the layer with an $R_x$ rotation on every qubit [Fig.~\ref{fig:qaoa_example_background}(b)].  
The $R_z$ and $R_x$ angles are tunable parameters, and a classical optimizer iteratively updates them to minimize the expected value of the Ising Hamiltonian measured on the quantum device.  
QAOA is regarded as one of the most promising approaches for demonstrating quantum utility on EFT systems.

\subsection{Prior Work Limitations}
Secure Quantum Computing (SQC) is an emerging field with numerous recent proposals, yet none offer practical solutions for the EFT era.

\subsubsection{Blind Quantum Computing (BQC)}
Blind Quantum Computing (BQC) enables a client to outsource a quantum computation to a remote quantum server while keeping the structure, inputs, and outputs of the computation hidden from the server~\cite{broadbent2009universal,li2021blind,morimae2013blind,fitzsimons2017private,fitzsimons2017unconditionally,childs2001secure}.  
These protocols provide strong cryptographic guarantees that the server learns nothing beyond the size of the computation.  
Most BQC protocols assume that the client possesses a small but trusted quantum device capable of preparing or measuring quantum states.  
They further assume the existence of a quantum communication channel between the client and server.  
However, large-scale, reliable quantum networks are not yet available, which makes this class of BQC protocols difficult to deploy in the EFT era.

Recent work has proposed alternative BQC protocols that remove the requirement for client-side quantum hardware by relying entirely on classical clients~\cite{reichardt2013classical,huang2017experimental,mckague2013interactive,gheorghiu2017rigidity,gheorghiu2015robustness,fitzsimons2017private}.  
These protocols achieve blindness by distributing trust across multiple non-communicating quantum servers.  
Security is preserved only if the servers are guaranteed to be isolated and unable to share information during the protocol execution.  
In practice, however, enforcing permanent physical or network isolation between untrusted quantum servers is nontrivial.

\subsubsection{Quantum Homomorphic Encryption (QHE)}
QHE extends the concept of homomorphic encryption to quantum data, allowing computations to be performed directly on encrypted quantum states~\cite{broadbent2015quantum,fisher2014quantum,broadbent2015delegating,dulek2016quantum,ouyang2018quantum,tan2016quantum,yu2014limitations,liang2013symmetric}.  
Unlike BQC, QHE eliminates the need for interactive communication between the client and server during computation.  
However, fully secure QHE requires exponential computational overhead, making it impractical for deployment in the EFT era, as tolerating such overhead would demand extensive QEC, and even with full QEC, sustaining exponential cost would remain highly challenging.

\subsubsection{Hardware Architecture for Trusted Quantum Execution} \label{sec:hardware_arch_trusted}
A recent work~\cite{trochatos2023hardware,trochatos2023quantum} proposes creating a trusted execution environment inside the dilution refrigerator of superconducting quantum devices, isolated from the outer region and inaccessible to the server.  
In this design, random decoy gates or pulses are inserted client-side after compilation and attenuated within the trusted region before execution.  
The regular structure of QAOA circuits further undermines this method, as each problem-graph edge maps to a fixed $R_z$--CNOT--CNOT pattern repeated with identical parameters across a layer, making decoys easy to detect.  
As QAOA is the leading candidate for quantum Ising solvers and its circuits are inherently repetitive, this technique cannot mitigate its privacy risks unless future algorithms emerge with less predictable structure.  
Finally, the approach was reasonable in the NISQ era, where machines operated directly on physical qubits and gates.  
As systems transition to EFT with logical operations under modest QEC, inserting and cancelling decoy logical operations becomes prohibitively complex and ultimately infeasible.

\subsection{Goal of this Paper}
\begin{figure}[t]
    \centering        
	\includegraphics[width=\columnwidth]{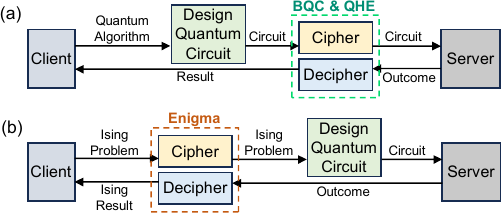}
    \caption{        
        Comparison of operational scope: (a) prior work applies obfuscation after generating the quantum program, 
        whereas (b) Enigma performs obfuscation before program generation at the application layer.
        }    
    \label{fig:prior_work_comparison}
\end{figure}

Previous SQC techniques aim to protect arbitrary quantum programs, requiring obfuscations after circuit generation or compilation, as shown in Figure~\ref{fig:prior_work_comparison}(a).  
While this approach maintains algorithm-level generality, it creates major deployment barriers, particularly in the EFT era.  
Many methods depend on technologies not yet available at scale, such as quantum networks or full-scale QEC, or are tied to a specific qubit technology.  
They are also designed primarily for the circuit-based model of quantum computing and cannot be directly applied to alternative paradigms such as measurement-based~\cite{raussendorf2001one}, adiabatic~\cite{albash2018adiabatic}, or emerging models such as fusion-based~\cite{bartolucci2023fusion} quantum computing.  

In this work, \emph{we introduce the notion of application-specific SQC and propose bringing obfuscation to the application layer}.  
By focusing on a single high-impact application, we strike a deliberate balance between generality and readiness for deployment.  
This paper presents \emph{Enigma}, a privacy-preserving framework that obfuscates Ising-model problem instances before generating quantum programs, as shown in Figure~\ref{fig:prior_work_comparison}(b).  

\vspace{0.1in}
\begin{tcolorbox}[colback=green!12]
By moving obfuscation to the application layer, Enigma is agnostic to (a) the quantum algorithm, enabling use beyond QAOA, (b) the computing model, supporting adiabatic, measurement-based, and emerging paradigms such as fusion-based computing, (c) the qubit technology, and (d) the type of accelerator, including quantum annealers.  
\end{tcolorbox}

\newpage
\section{Threat Model}

The goal of Enigma is to protect sensitive information about a user’s quantum optimization problem when outsourcing computation to an untrusted server. 
This includes the number of variables, problem coefficients, graph structure, and the solution itself. 
The only secret in our model is the randomness used during obfuscation; everything else, including the details of Enigma's algorithms, is public and visible to the server.

We assume a conservatively strong adversary: the server, or a malicious insider, has full access to all submitted circuits and aims to deobfuscate the original problem or learn any information about the underlying optimization problme. 
The adversary possesses a large database of canonical quantum optimization problems (potentially collected from various clients) and can apply Enigma to its own database, producing obfuscated versions of known instances to perform statistical inference or repository-matching attacks against user-submitted circuits. We assume the circuit’s raw measurement outcomes reveal nothing meaningful without knowledge of the original problem graph.

Given the structured nature of quantum optimization circuits, a server may still recognize that a submitted circuit corresponds to optimization.  
This is acceptable, as Enigma is not intended to conceal the circuit type but rather to hide the specific problem instance.

We assume the server does not tamper with outputs, as incorrect results can be detected via solution quality checks, and we consider other active attacks out of scope.


\section{Application-Layer Obfuscations for Ising-Model Problem Instances}
\label{sec:obfuscations}

In this section, we present a suite of application-specific SQC techniques that obfuscate Ising problem instances at the application layer, before circuit generation.  
All schemes use the same Ising model input/output format, enabling plug-and-play composition in any order.

\subsection{ValueGuard: Masking Coefficients and Solutions}  
\label{sec:valueguard}

\begin{figure}[t]
    \centering        
	\includegraphics[width=0.80\columnwidth]{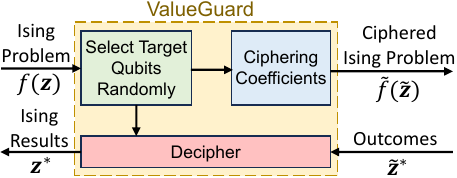}	
    \caption{
        Overview of ValueGuard
        }    
    \label{fig:e1_overview}
\end{figure}

We introduce \emph{ValueGuard}, an obfuscation scheme that protects Ising-model problem instances by hiding both the coefficients and the solution bitstrings.  
ValueGuard randomly selects a subset of spins $\mathbf{z} \in \{-1, +1\}$ and flips the signs of all coefficients associated with those spins.
It then multiplies all coefficients by a random positive scaling factor to further obscure their true values.
The sign flips correspond to bit flips in the binary domain, which means the outcomes of the selected spins are deliberately inverted.
The server observes and solves a transformed problem whose optimal solution is systematically scrambled.
Only the client, who knows exactly which spins were flipped, can reverse these transformations and recover the correct original solution from the server’s output.
Figure~\ref{fig:e1_overview} illustrates this process.


\begin{figure*}[t]
    \centering        
    \includegraphics[width=\textwidth]{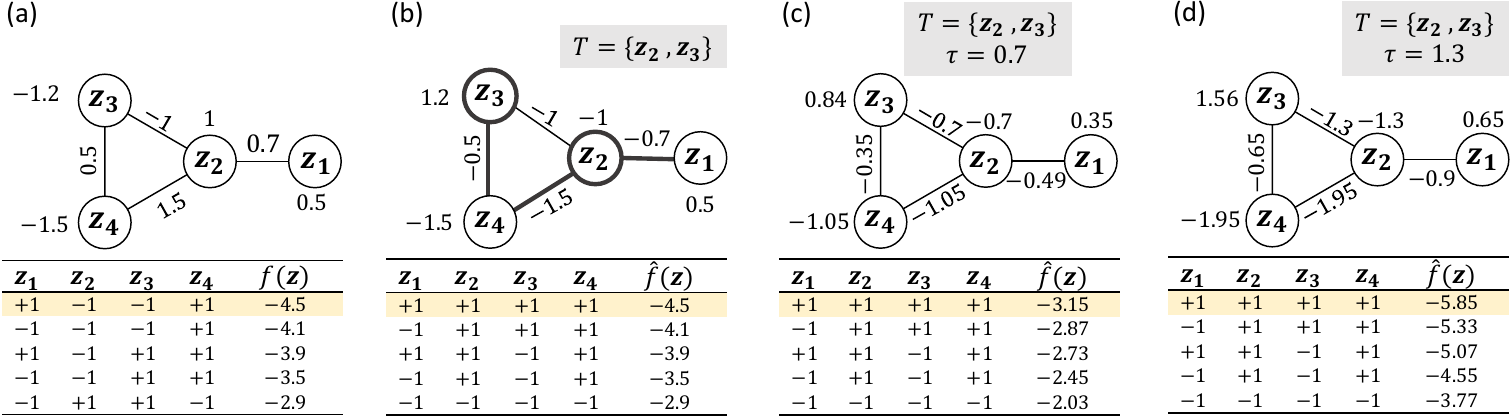}	
    \caption{                
A four-node Ising problem with a cost table of the top five of 16 configurations, ordered from lowest (top) to highest objective value, with the global minimum highlighted.   
(a) Original Ising instance.  
(b) After flipping spins of $\mathbf{z}_1$ and $\mathbf{z}_2$.  
(c) Scaling all coefficients in (b) by $\tau < 1$, reducing energy gaps.  
(d) Scaling all coefficients in (b) by $\tau > 1$, increasing energy gaps.
        }    
    \label{fig:e1_encrypt_coeffs}
\end{figure*}

\subsubsection{Target Spins: Decrypting Key}
ValueGuard begins by selecting a set of \emph{target spins}, denoted by $T$.  
These spins are chosen uniformly and independently at random, so the size and composition of $T$ vary with each instance.  
The list $T$ serves as the Decrypting key and must be kept confidential, since anyone with access to it can invert the transformations and recover the original solution distribution.

\subsubsection{Encrypting Coefficient Signs and Inducing Spin Flips}
For each target spin $\mathbf{z}_i \in T$, ValueGuard negates its linear coefficient: $\hat{\mathbf{h}}_i = -\mathbf{h}_i$.  
For quadratic terms, if exactly one of $\mathbf{z}_i$ or $\mathbf{z}_j$ is in $T$, the coefficient is negated: $\hat{J}_{ij} = -J_{ij}$.  
As illustrated in Fig.~\ref{fig:e1_encrypt_coeffs}(b), these sign changes flip the values of target spins in the ground-state configuration without altering the Hamiltonian’s minimum energy.  
Spin flips preserve the Ising energy spectrum while permuting spin assignments across energy levels, concealing both the coefficient signs and the true spin values.  

\subsubsection{Stretching: Obfuscating Coefficient Magnitudes} \label{sec:e2_Stretching}
Multiplying the objective $f$ by any positive scalar $\tau > 0$ rescales energies but preserves the ordering of $f$'s values and the global minimum.  
Leveraging this, ValueGuard multiplies all coefficients, $\hat{\mathbf{h}}$ and $\hat{J}$, by a randomly generated positive scalar $\tau$ to hide coefficient magnitudes.  

For $0 < \tau < 1$, $f$ is contracted (Fig.~\ref{fig:e1_encrypt_coeffs}(c)).  
For $\tau > 1$, $f$ is stretched (Fig.~\ref{fig:e1_encrypt_coeffs}(d)).  
For $\tau = 1$, $f$ is unchanged (Fig.~\ref{fig:e1_encrypt_coeffs}(b)).  
By default, ValueGuard samples $\tau$ from a log-uniform distribution, giving equal probability to small and large scaling factors.  
We first draw a random value $u$ uniformly from the range $[-2, 2]$, then set  
\begin{equation}
\tau = 10^{u},
\label{eqn:e2_stretching}
\end{equation}
which produces $\tau$ values between $10^{-2}$ and $10^{2}$.  
This range balances effective obfuscation with numerical stability.

\subsubsection{Decrypting Outcomes} \label{sec:e1_decoding}
After the server solves the obfuscated Ising instance, the client uses the target qubit list $T$ to recover the original solution.  
This is done by flipping the spins (or equivalently, bits) of all qubits in $T$, restoring the correct bitstring without changing its probability.  
$T$ is the only information needed for decoding, so the client keeps it secret; the coefficients do not need to be recovered, as the client already possesses the original problem.

\subsection{StructureCamouflage: Adding Decoy Variables}  
\label{sec:decoy_vars}

\begin{figure}[h]
    \centering        
	\includegraphics[width=1\columnwidth]{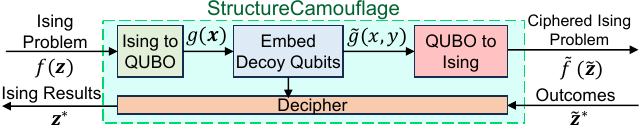}	
    \caption{
        Overview of StructureCamouflage
    }    
    \label{fig:e2_overview}
\end{figure}

We introduce \emph{StructureCamouflage} (Fig~\ref{fig:e2_overview}), an obfuscation scheme for Ising-model problem instances that conceals both problem variables and graph structure by embedding \emph{decoy} nodes and edges that are indistinguishable from the original ones.  
Given an Ising instance, StructureCamouflage:
(1) converts it to the binary domain,
(2) inserts decoy variables,
(3) adds edges between primary and decoy variables, as well as among decoys,
(4) assigns randomized weights to these edges, and
(5) converts the obfuscated instance back to the Ising domain.  
When Decrypting, the client simply discards all decoy variables from the server’s returned solution.

\begin{figure*}[b]
    \centering        
	\includegraphics[width=0.7\textwidth]{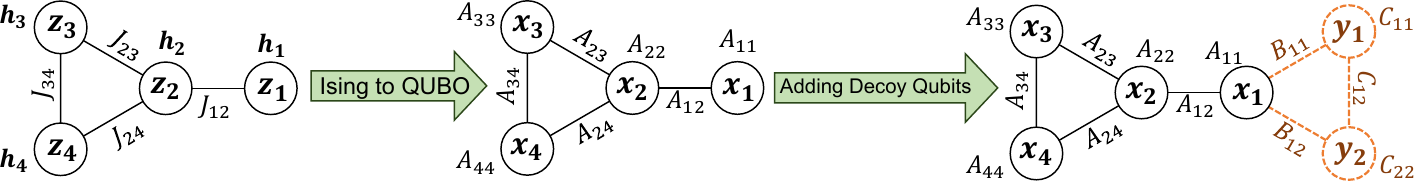}
    \caption{
        StructureCamouflage converts the input Ising model to QUBO form to introduce decoy nodes ($\mathbf{y}$) and edges.
    }    
    \label{fig:e2_example}
\end{figure*}

\subsubsection{Transforming Ising to QUBO}

Any Ising model can be transformed into an equivalent \emph{Quadratic Unconstrained Binary Optimization} (QUBO) formulation where spin variables $\mathbf{z}_i \in \{-1,+1\}$ are replaced by binary variables $\mathbf{x}_i \in \{0,1\}$~\cite{boettcher2019analysis,lucas2014ising,ayanzadeh2024skipper,ayanzadeh2020reinforcement,ayanzadeh2022quantum} (see Appendix~\ref{apx:ising_qubo}).   
In the binary domain, the objective function is expressed as  
\begin{equation}
	g(\mathbf{x}) =
    \sum_{i=1}^{n} A_{ii} \mathbf{x}_i +
		\sum_{i=1}^{n} \sum_{j=i+1}^{n} A_{ij} \mathbf{x}_i \mathbf{x}_j, 
        \label{eqn:qubo}
\end{equation}
where $A$ is the coefficient matrix with diagonal entries as linear and off-diagonal entries as quadratic coefficients.

StructureCamouflage first converts the input Ising instance to its QUBO form (Fig.~\ref{fig:e2_example}).  
In the binary domain, the QUBO objective admits a matrix representation.  
We exploit the algebraic properties of this form to guarantee recovery of the original problem’s global minimum from the obfuscated solution (Section~\ref{sec:successful_recovery}, Appendix~\ref{apx:decoy_var}).

\subsubsection{Adding Decoy Nodes and Edges}
To obfuscate the problem graph topology, StructureCamouflage introduces \emph{decoy nodes} $\mathbf{y}_k \in \{0,1\}$ (Fig.~\ref{fig:e2_example}).  
To make these nodes indistinguishable from real nodes $\mathbf{x}_i$,  
\emph{decoy edges} are added between real and decoy nodes, with coefficients $B_{ik}$ for edges between $x_i$ and $y_k$,  
and between pairs of decoy nodes, with coefficients $C_{kl}$ for edges between $y_k$ and $y_l$, and $C_{kk}$ as the linear coefficient of $y_k$.

\subsubsection{Generating Decoy Coefficients with Guaranteed Recovery}  
\label{sec:successful_recovery}

Adding decoy nodes and edges yields:
\noindent
\begin{equation}    
    \SMALL
    \tilde{g}(\mathbf{x},\mathbf{y}) = g(\mathbf{x}) + \sum_{k=1}^{m} C_{kk}\mathbf{y}_k + \sum_{i=1}^{n} \sum_{k=1}^{m} B_{ik}\mathbf{x}_i \mathbf{y}_k \nonumber + \sum_{k=1}^{m} \sum_{l=k+1}^{m} C_{kl}\mathbf{y}_k \mathbf{y}_l,
    \label{eqn:qubo_aug}
\end{equation}
\noindent
Note that edges are not added between real nodes, as such modifications would alter the original problem’s objective function and prevent guaranteed recovery of its global optimum from the obfuscated solution.  
StructureCamouflage adds decoy edges using four strategies: random, low-degree preference, high-degree preference, or full connectivity.

Protecting the privacy of quantum optimization workloads is critical, but the transformed problem must also remain solvable to fulfill the original optimization objective.  
The coefficients $B$ and $C$ cannot be arbitrary without risking loss of recoverability.  
\textbf{A sufficient condition for guaranteeing recovery of the original problem’s global optimum from the obfuscated solution is for all entries of $B$ and $C$ to be non-negative (see Appendix~\ref{apx:decoy_var} for proof).}

\begin{figure}[b]
    \centering        
	\includegraphics[width=\columnwidth]{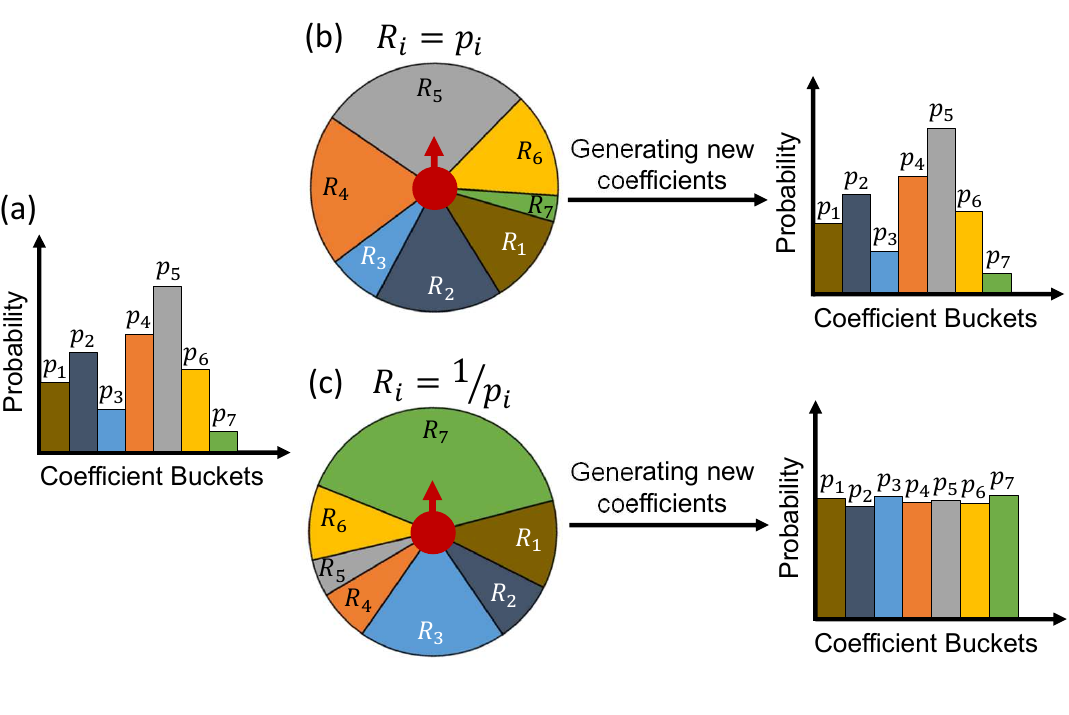}
    \caption{
        (a) Original coefficient distribution.  
        (b) Sampling decoy edge weights proportional to the existing distribution preserves its profile.  
        (c) Sampling inversely proportional to the distribution yields a more uniform profile.
    }    
    \label{fig:e2_roulette_wheel}
\end{figure}

While any non-negative $B$ and $C$ entries are sufficient for successful recovery, care is needed when assigning decoy edge weights.  
A simple approach is to generate coefficients randomly, for example by sampling from a uniform or normal distribution.  
However, if the number of decoy coefficients is small relative to the number of primary coefficients and the primary coefficients follow a distinct distribution, an adversary may identify some decoy edges and, by extension, some decoy nodes, reducing the effectiveness of obfuscation.  
To mitigate coefficient-distribution leakage, StructureCamouflage assigns decoy weights using an inverse-distribution roulette-wheel sampling method (Fig.~\ref{fig:e2_roulette_wheel}) inspired by evolutionary algorithms~\cite{lipowski2012roulette,shukla2015comparative,eiben2015introduction}.  
The absolute values of the original coefficients are first binned into $b$ intervals (default $b=10$), and each bin is selected with probability $p_i$ proportional to its frequency.  
Two modes are supported:  
(a)~$R_i = p_i$ preserves the original distribution, and  
(b)~$R_i = 1/p_i$ flattens the distribution toward uniformity (default).  
This flattening maximizes entropy and makes decoy and real coefficients statistically indistinguishable, thereby concealing the original coefficient distribution.

\subsubsection{Decrypting Outcomes}
The client recovers the original solution by removing all decoy variable assignments from the server’s returned bitstrings.  
Because only the client knows which variables are decoys, this step is trivial for the client yet infeasible for the server without that information.

\subsection{TopologyTrimmer: Stochastic Node Removal}  
\label{sec:node_removal}

\begin{figure}[b]
    \centering        
	\includegraphics[width=\columnwidth]{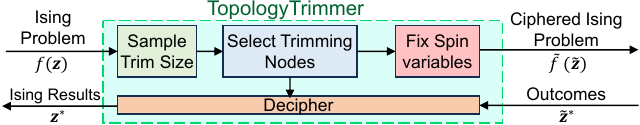}	
    \caption{
        Overview of TopologyTrimmer
    }    
    \label{fig:fq_overview}
\end{figure}

Inspired by prior work~\cite{ayanzadeh2023frozenqubits}, we introduce \emph{TopologyTrimmer}, a graph-aware obfuscation scheme that perturbs an Ising problem’s topology by pruning nodes and their incident edges (Fig.~\ref{fig:fq_overview}).  
Given an input, it produces multiple smaller Ising instances with identical structure, making them difficult for an adversary to distinguish.  
Unlike prior work~\cite{ayanzadeh2023frozenqubits}, which always removes hotspots, both the number and identity of trimmed nodes are stochastic here, so repeated application generates subproblems with varying sizes and topologies, significantly complicating recovery of the original problem.  
For decryption, the client restores the trimmed variables in each solution using their fixed values.

\subsubsection{Selecting Trimmed Variables}
The number of trimmed variables $o$ is drawn uniformly from $\{0,1,\dots,10\}$, introducing nondeterminism even for a fixed input.  
This randomness allows each run to remove a different number of nodes, including possibly none.  
The user may override this behavior and set $o$ explicitly.

\subsubsection{Selecting Trimmed Nodes}  
The choice of trimmed nodes in TopologyTrimmer is both structure-aware and stochastic, supporting the following strategies:  
\emph{Hotspot} — $o$ highest-degree nodes;  
\emph{Uniform} — all nodes equally likely;  
\emph{High-Degree-Biased} — probability proportional to degree;  
\emph{Low-Degree-Biased} — probability inversely proportional to degree;  
\emph{Hybrid} — half Hotspot, half High-Degree-Biased.  
These strategies balance structural awareness with stochastic variation across runs.

\subsubsection{Trimming Nodes}
TopologyTrimmer applies the node-freezing strategy from FrozenQubits~\cite{ayanzadeh2023frozenqubits}.  
Starting from the original Ising instance, each of the $o$ trimmed nodes is processed sequentially by duplicating every current instance and fixing the node to $-1$ and $+1$ in the two copies.  
This process generates $2^{o}$ subproblems, all of which are solved to preserve optimization quality (SEE Appendix~\ref{apx:FQ}).  

\subsubsection{Decrypting Outcomes}
The client reconstructs the original solution by appending the removed variables and their fixed values to the server’s output bitstrings.  
Because only the client knows which variables were removed and how they were fixed, this step is straightforward for the client and infeasible for the server.

\newpage
\section{Enigma: Adaptive Obfuscation for Secure Quantum Optimization}
\label{sec:enigma}

Building on the obfuscation schemes introduced in Section~\ref{sec:obfuscations}, we present \emph{Enigma}, an adaptive, graph-aware framework for secure quantum optimization.

\subsection{Operation Model of Enigma}
\begin{figure}[h]
    \centering        
	\includegraphics[width=\columnwidth]{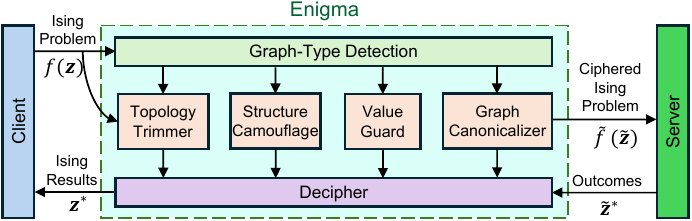}	
    \caption{
        Enigma Overview        
    }    
    \label{fig:enigma_overview}
\end{figure}

Enigma integrates multiple obfuscation modules with a canonicalization stage, producing solver-ready Ising instances whose solutions can be efficiently Decrypted by the client.  
\emph{The obfuscation schemes introduced in Section~\ref{sec:obfuscations} all operate directly on Ising models, enabling a composable, order-agnostic workflow where modules can be applied in any sequence and even repeated.}  
In this work, we design a deliberate, performance-conscious pipeline that maximizes security while preserving solution quality (Fig~\ref{fig:enigma_overview}).  
Enigma first detects the problem’s graph type and adaptively configures obfuscation parameters, reflecting its graph-aware design.
The chosen sequence is \emph{TopologyTrimmer} $\rightarrow$ \emph{StructureCamouflage} $\rightarrow$ \emph{ValueGuard}.  
To remove any residual footprint from earlier stages, we then apply \emph{GraphCanonicalizer} as a cleanup step (Section~\ref{sec:cleanup}).  

After cleanup, the obfuscated Ising instance(s) are sent to the solver.  
A given original problem may yield multiple obfuscated instances when TopologyTrimmer produces subproblems.  
The client runs all instances, collects their solutions, and applies Decrypting in reverse order of obfuscation to reconstruct the original problem’s solution.

\subsection{Graph-Type Detection}
\label{sec:graph_type}  
Identifying regular graphs is trivial: all nodes have equal degree, with $n-1$ in the fully connected case.  
To distinguish Erdős–Rényi (ER) from power-law or Barabási–Albert (BA) graphs, we propose the following heuristic:
\[
\max(\mathbf{d}) > 1.9 \,\mathrm{mean}(\mathbf{d}) \Rightarrow \text{BA}, \quad \text{else ER},
\]  
where $\mathbf{d}$ is the node degree list.  
For $n>50$, we observe this heuristic yields $100\%$ accuracy.


\subsection{Obfuscation Parameterization}  
\label{sec:ob_parameterization}

Enigma tailors its obfuscation parameters to the detected graph type, balancing obfuscation strength against system overhead.  
Table~\ref{tbl:enigma_setting} lists the critical settings for each graph category, reflecting this trade-off.


\begin{table}[t]
\centering
\small
\caption{
    Obfuscation settings by graph type in Enigma.   
}
\begin{tabular}{lccc}
\toprule
Topology & Trim Strategy & Decoy Nodes (\%) & Edge Attach \\
\midrule
SK   & Random         & 1--7.5        & Full \\
Reg  & Random         & 1--5       & LowDeg \\
BA   & Hybrid         & 5--15      & LowDeg \\
ER   & Hybrid   & 5--25      & LowDeg \\
\bottomrule
\end{tabular}
\label{tbl:enigma_setting}
\end{table}

\subsection{GraphCanonicalizer}
\label{sec:cleanup}

The \emph{GraphCanonicalizer} module is the final stage before sending the obfuscated Ising instance to the solver.  
Its role is to \emph{remove residual footprints from prior obfuscation steps}, eliminating structural artifacts that might reveal how the instance was transformed.  
It enforces a clean and  consistent representation through the following actions:

\begin{itemize}
    \item \textbf{Removal of zero-degree nodes:} Nodes without incident edges are removed, as they do not affect the optimization objective and could trivially reveal primary variables.
    
    \item \textbf{Edge normalization ($i<j$ in $J$):} All quadratic terms $J_{i,j}$ are stored with $i<j$.  
    Without normalization, consistent lower- or upper-triangular patterns from earlier toolchains could leak which edges are original versus decoys.  
    Normalization removes this bias.
    
    \item \textbf{Variable re-indexing and shuffling:} Variable indices are remapped to a contiguous range starting at zero and then shuffled.  
    This both fills gaps left by node removals and prevents an adversary from inferring that highest-numbered variables are newly added decoys.  
    The client retains the mapping for trivial decoding; without it, an adversary faces a non-trivial reconstruction challenge.
\end{itemize}

These steps preserve the optimization semantics while eliminating any statistical or structural signatures left by earlier obfuscations.

\subsection{Decrypting Outcomes}
Enigma’s Decrypting reverses the obfuscations in the opposite order of their application.  
\emph{Since the client owns the original problem, only the solution needs reconstruction, not the original graph.}  
First, variable names are restored to their original indices to undo re-indexing and shuffling.  
For zero-degree nodes removed during cleanup, solutions are expanded by sequentially reintroducing each node with both fixed values $-1$ and $+1$, doubling the solution count per node and yielding $2^{\tilde{o}}$ solutions when $\tilde{o}$ nodes are reintroduced.  
Next, ValueGuard decoding flips the spins of variables in its target list.  
StructureCamouflage decoding then removes decoy variables.  
Finally, for each solution of every TopologyTrimmer subproblem, trimmed variables are restored with their fixed values.  
Among the global optima of all subproblems, the best one is returned as the solution to the original Ising problem.

\section{Enigma Analysis and Results}

\subsection{Methodology}

\subsubsection{Benchmarks}
We evaluate Enigma on five representative graph families: fully connected (SK), 3-regular, Erdős–Rényi (ER), and Barabási–Albert (BA)~\cite{albert2005scale,barabasi1999emergence} with attachment factors of 1 (BA-1) and 2 (BA-2).  
BA graphs approximate the power-law structures commonly observed in real-world networks~\cite{clauset2016colorado,broido2019scale,agler2016microbial,barabasi1999emergence,barabasi2000scale,gray2018super,kim2022sparsity,lusseau2003emergent}, while SK, regular, and ER graphs serve as canonical benchmarks.  
All graphs are generated using NetworkX random graph generators with edge weights drawn uniformly from $[-1,1]$, following common practice in quantum optimization benchmarking~\cite{alam2020efficient,ayanzadeh2022equal,ayanzadeh2021multi,ayanzadeh2023enhancing}.  
Graph sizes are sampled uniformly between 500 and 1000 nodes, consistent with prior work~\cite{guerreschi2019qaoa} indicating that several hundred variables are necessary to capture meaningful problem hardness.

\subsubsection{Experimental Setup for Security Analysis}
\label{sec:method:security}

\paragraph{Adversaries' graph dataset.}
For our security evaluation, we assume an adversary with access to a database of \(G{=}10{,}000\) candidate original graphs and their corresponding obfuscated graphs produced by Enigma.  
The database is class-balanced across five graph types: \texttt{BA1}, \texttt{BA2}, \texttt{ER}, \texttt{3-REGULAR}, and \texttt{SK}.  
For training, each original graph is paired with its obfuscated counterpart.  
For validation, we create queries from a disjoint set of 100 original graphs per type, each independently obfuscated.



 
 
 
 
 

\paragraph{Deobfuscation models.}
We evaluate Enigma against seven state-of-the-art graph de-anonymization algorithms drawn from the literature.  
The selected models include both feature-based and structure-based approaches to capture a broad range of adversarial strategies.  

\vspace{0.05in}
\noindent
\textbf{Baselines.}
We evaluate six \emph{feature-based} baselines that operate on 70-dimensional vectors summarizing each QUBO instance using simple statistics over the Ising/QUBO coefficients.  
These include counts of diagonal and off-diagonal terms, sign counts, and basic distributional summaries.  
The summaries are fixed and not learned from data.  
The baselines are as follows.  
\begin{itemize}[leftmargin=*] 
    \item \emph{Direct cosine similarity}: a standard retrieval baseline in re-identification and entity matching~\cite{Christen2012,Mudgal2018}.  
    \item \emph{$k$-NN (features)}: a non-parametric linkage baseline using the same standardized summaries~\cite{Christen2012,Mudgal2018}.  
    \item \emph{Random forest}: trained to classify concatenated \(\langle\)obfuscated, original\(\rangle\) feature pairs as match or non-match~\cite{Christen2012,Mudgal2018}.  
    \item \emph{Siamese neural network}: learns embeddings that place true pairs closer together than non-matches~\cite{Schroff2015,Hoffer2015,Khosla2020}.  
    \item \emph{Contrastive-learning model}: directly optimizes pairwise distances with a margin between matched and mismatched pairs~\cite{Khosla2020,Bellet2013}.  
    \item \emph{Feature-weighted similarity}: learns per-feature weights to improve separation, providing an interpretable instance of discriminative metric learning~\cite{Bellet2013}.  
\end{itemize}

\noindent
Additionally, we include one \emph{structure-only, seedless} baseline: the \textbf{Seedless FGW matcher}.  
This method aligns each query to every original graph in the reference set using fused Gromov–Wasserstein optimal transport and relies on lightweight NetLSD heat-trace sketches and Weisfeiler–Lehman node patterns to capture both global and local structure~\cite{Tsitsulin2018,Shervashidze2011,Vayer2019,Flamary2021}.  
\textbf{We emphasize the feature-based baselines because, in early full-dataset experiments on our QUBO/Ising graphs and obfuscations, they consistently outperformed structure-aware and GNN-based models under the same protocol.}  

\vspace{0.05in}
\noindent
\textbf{Metric.}
We report \emph{top-5 accuracy} as the ranking-based measure of attacker success, following prior graph de-anonymization work~\cite{Ji2015}.



\subsubsection{Experimental Setup for Overhead Analysis}

\paragraph{Hardware Platform.}
We target the EFT era and assume parameters achievable in the near term.  
Specifically, we consider a quantum device with grid connectivity and a physical error rate of $10^{-4}$.  
Using the surface code with distances between five and seven yields logical error rates in the range $10^{-6}$–$10^{-9}$, which we expect to be realistic in this era.  
Real-time decoding has already been demonstrated at code distances up to 9~\cite{vittal2023astrea} and 11~\cite{alavisamani2024promatch}, reinforcing this assumption.  

\paragraph{Ising Solver.}
As a proof of concept, we evaluate Enigma on QAOA.  
While Enigma is agnostic to algorithm, we chose QAOA as it is currently the most competitive quantum optimization method.  
All overhead analysis is therefore based on Enigma’s impact on QAOA programs.  

\paragraph{Metrics}
We evaluate three overhead types.  
For program fidelity, we use success probability, reflecting logical error rates achievable with modest QEC.  
For graph size, we report node and edge counts, average degree, and density.  
For QEC overhead, we measure Enigma’s impact by the change in total $T$-gate count.  
In leading QEC schemes such as the surface code, Clifford gates are implemented transversally, while non-Clifford $T$ gates require costly magic state distillation~\cite{bravyi2005universal,dangwal2025variational,hao2025reducing}.  
Since $T$ gates dominate resource costs, their count reflects Enigma’s QEC overhead.

\begin{figure}[b]
    \centering        
	\includegraphics[width=\linewidth]{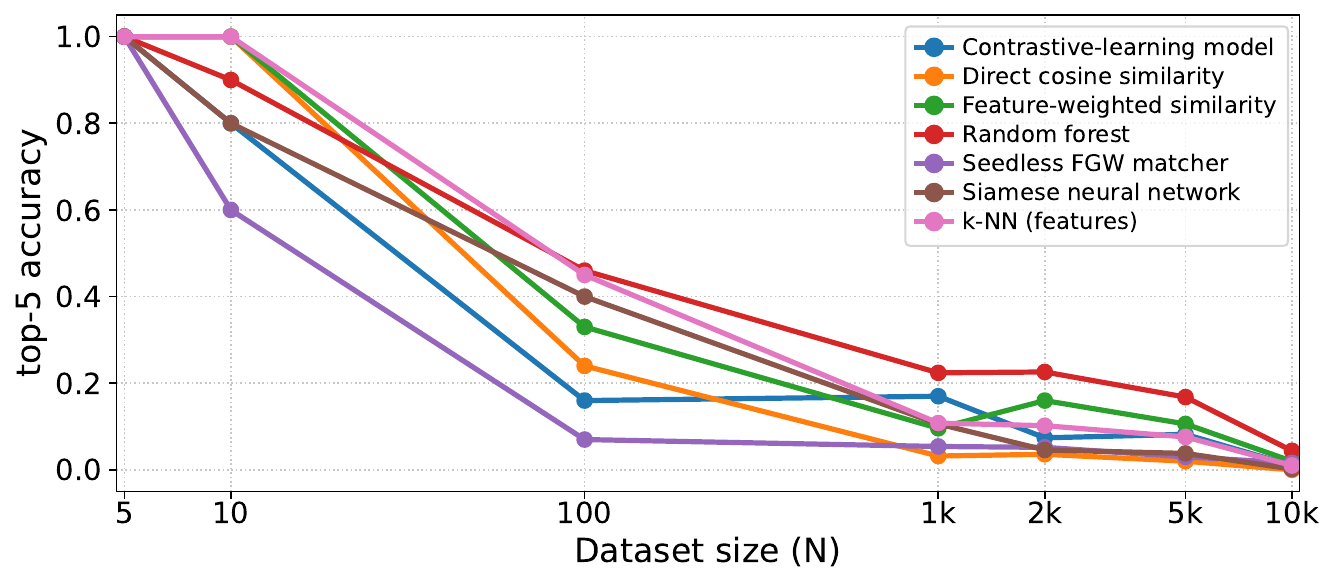}	
    \caption{
        The results of security analysis. All models fail to achieve a meaningful top-5 accuracy for our dataset with 5k graphs. 
        }    
    \label{fig:SQC_results}
\end{figure}

\subsection{Security Analysis and Results}
This section present the results of our empirical security evaluation.
In order to show the effectiveness of obfuscation of Enigma, we use several state of the art graph deanonymization models as described in Section~\ref{sec:method:security}.

\paragraph{All the models fail to deanonymize Enigma.}
The goal of the adversary is to use the best statistical model available to de-anonymize an obfuscated graph produced by Enigma.  
Each graph in the test set corresponds to an obfuscated version of one of the graphs in the adversary’s dataset.  
The adversary trains models on this dataset and, given an obfuscated test sample, attempts to attribute it to the correct original graph.  

With a dataset of 10{,}000 graphs, the best-performing adversary model achieves only 4.4\% top-5 accuracy.  
This means that in just 4.4\% of test cases, the correct original graph appears within the model’s top five guesses, demonstrating the strength of Enigma’s obfuscation.

\paragraph{Sensitivity to Dataset Size.}
To confirm that this result stems from Enigma’s protection rather than model weakness, we re-evaluate the same adversaries on an easier task by reducing the output space, i.e., the number of possible original graphs.  
As expected, performance improves as the output space shrinks, showing that the models can de-anonymize effectively when the task is simpler.  
For the learned baselines (Random Forest, Siamese, Contrastive, and Feature-Weighted Similarity), we also observe a modest improvement when moving from very small datasets to moderate ones, reflecting the benefit of additional training pairs.  
Beyond that point, as the output space grows, the increased complexity dominates and accuracy declines.

Figure~\ref{fig:SQC_results} shows the results of this experiment.  
We also find that model performance depends on the problem size, with different models excelling in different settings.  
Table~\ref{tab:security-best-overall} reports the best overall top-5 accuracy and the mean reciprocal rank (MRR) across dataset sizes.  

At dataset size $5$, all models achieve $1.00$ top-5 accuracy.  
At $10$, most remain near perfect.  
By $100$, several attacks reach $0.24$–$0.46$ top-5 accuracy.  
As size increases, accuracy drops sharply: the best at $1000$ is $0.224$ (\emph{Random Forest}); at $2000$, $0.226$ (tie between \emph{Random Forest} and \emph{$k$-NN (features)}); at $5000$, $0.168$ (\emph{$k$-NN (features)}); and at our primary size $10{,}000$, $0.044$ (\emph{Random Forest}).

\begin{table}[t]
\centering
\small
\begin{threeparttable}
\caption{Best overall \(\text{top-5 accuracy}\) and MRR per dataset size \(G\).}
\label{tab:security-best-overall}
\begin{tabular}{r@{\quad}l@{\quad}r@{\quad}r}
\toprule
\(G\) & Best model & \(\text{top-5 accuracy}\) & MRR \\
\midrule
\(5\)        & All models                 & \(1.000\) & \(1.000\) \\
\(10\)       & Multiple models\tnote{*}   & \(1.000\) & \(0.758\) \\
\(100\)      & Random Forest (features)   & \(0.460\) & \(0.282\) \\
\(1000\)     & Random Forest (features)   & \(0.224\) & \(0.170\) \\
\(2000\)     & Random Forest (features)   & \(0.226\) & \(0.160\) \\
\(5000\)     & Random Forest (features)   & \(0.168\) & \(0.112\) \\
\(10{,}000\) & Random Forest (features)   & \(0.044\) & \(0.043\) \\
\bottomrule
\end{tabular}
\begin{tablenotes}[flushleft]
\footnotesize
\item[*] Cosine (features), \(k\)-NN (features), and Weighted-Feat all achieve perfect top-5 accuracy.
\end{tablenotes}
\end{threeparttable}
\end{table}

\begin{figure*}[t]
    \centering        
	\includegraphics[width=\textwidth]{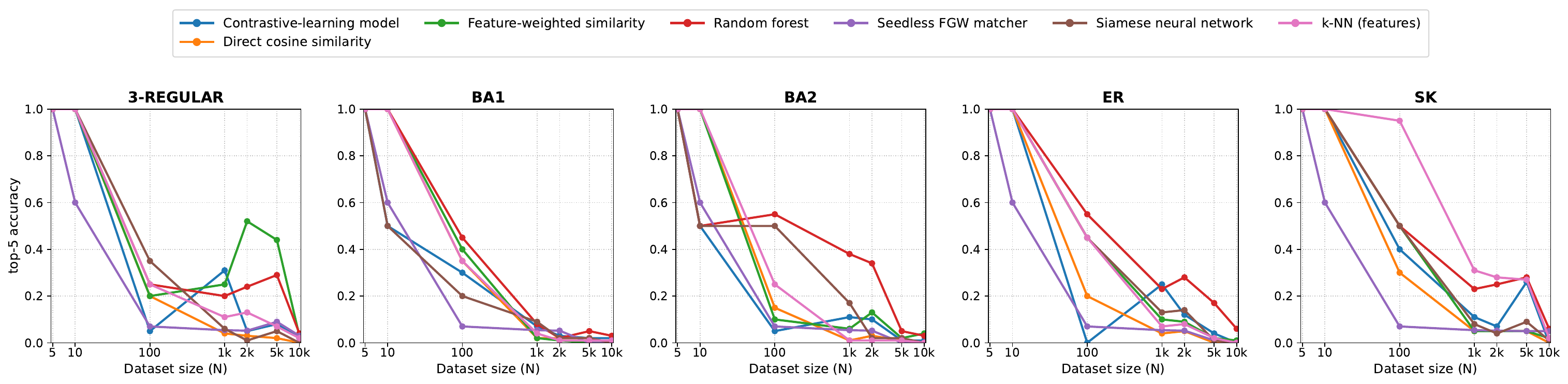}	
    \caption{
        Breakdown of accuracy of our deanonymization models for different types of graphs.
        }    
    \label{fig:SQC_impact_graph_type}
\end{figure*}

\paragraph{Which graph types are easier to re-identify?}
Performance varies by graph type and dataset size (Fig.~\ref{fig:SQC_impact_graph_type}).  
At size $100$, \texttt{sk} is the easiest to attack (e.g., \emph{$k$-NN (features)} achieves top-5 accuracy $0.95$), while \texttt{ba2} is inconsistent across models (ranging from $0.05$ with the \emph{Contrastive-learning model} up to $0.55$ with \emph{Random Forest}).  
At sizes $1000$–$2000$, \texttt{sk} remains the most vulnerable among the feature baselines (\emph{$k$-NN (features)}: $0.31 \rightarrow 0.28$).  
\emph{Random Forest} is next-best on \texttt{er} ($0.23 \rightarrow 0.28$) and also strong on \texttt{ba2} ($0.38 \rightarrow 0.34$).  
\emph{Feature-weighted similarity} peaks on \texttt{3-regular} ($0.52$ at size $2000$).  
The \emph{Seedless FGW matcher} remains consistently low (about $0.05$–$0.07$).  
By size $5000$, nontrivial signal persists (e.g., \emph{Random Forest}: \texttt{sk} $0.28$, \texttt{3-regular} $0.29$; \emph{$k$-NN (features)}: \texttt{sk} $0.27$).  
At size $10{,}000$, accuracies decline but remain materially above chance (top-5 chance is $0.0005$), with \emph{Random Forest} reaching about $0.06$ on \texttt{sk} and \texttt{er}.  
Overall, \texttt{sk} is consistently the easiest to re-identify, \texttt{3-regular} and \texttt{er} retain moderate signal at larger sizes, and \texttt{ba2} exhibits strong model dependence.  


\subsection{Enigma Overhead Analysis and Results}

\subsubsection{Impact on Problem Size}
\begin{figure}[h]
    \captionsetup[subfigure]{position=top} 
    \centering
    \subfloat[]{
		\includegraphics[width=0.47\columnwidth]{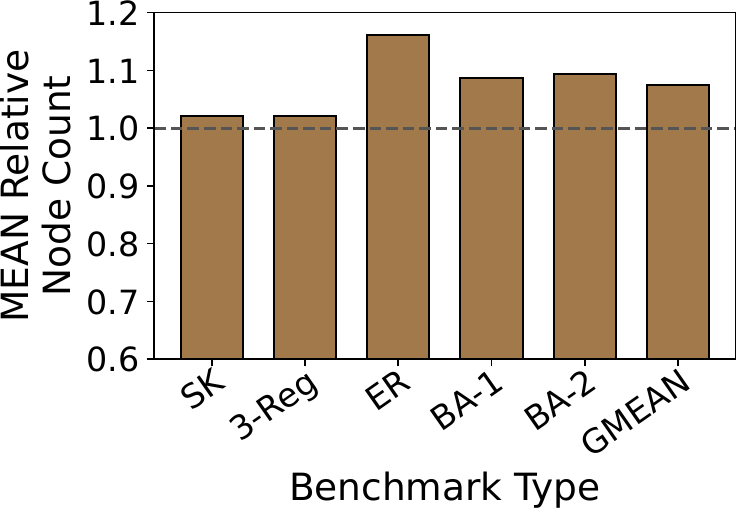}
	}
    \subfloat[]{	
        \includegraphics[width=0.47\columnwidth]{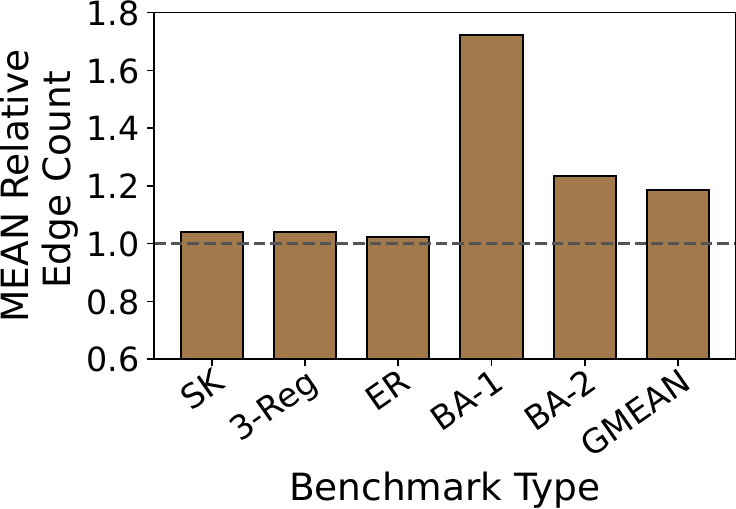}
	}
    \\
    \subfloat[]{	
        \includegraphics[width=0.47\columnwidth]{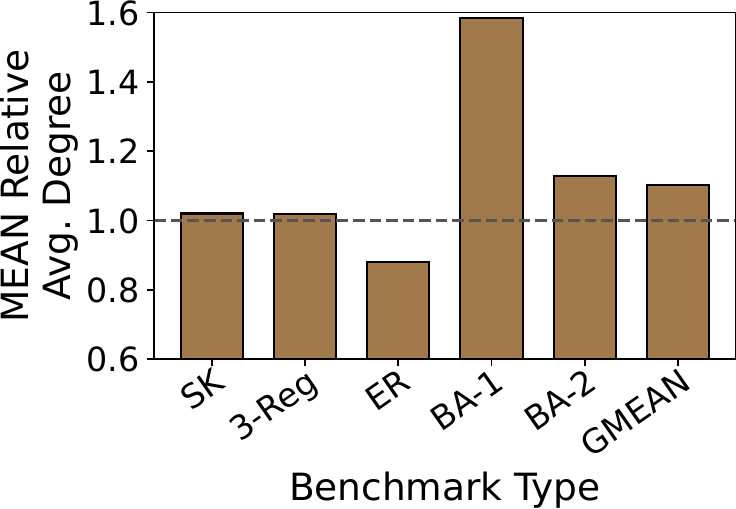}
	}
    \subfloat[]{	
        \includegraphics[width=0.47\columnwidth]{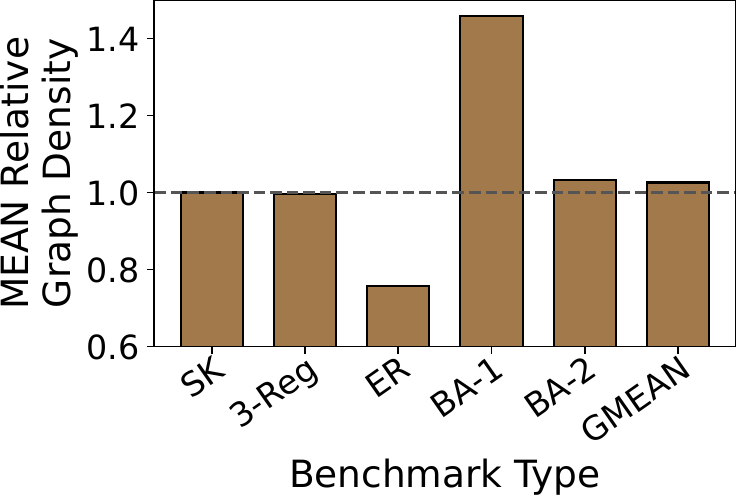}
	}
    \caption{
        Impact of Enigma on problem size across four metrics: 
        (a) relative node count;
        (b) relative edge count;  
        (c) relative average degree $\bar{d}=2E/N$; and   
        (d) relative density $D=2E/(N(N-1))$ (lower is better). 
    }    
    \label{fig:impact_graph}    
\end{figure}

We evaluate Enigma's impact on problem size across four metrics.  
As shown in Figure~\ref{fig:impact_graph}(a), Enigma increases node count by 1.07$\times$ on average (up to 1.16$\times$).  
As shown in Figure~\ref{fig:impact_graph}(b), it increases edge count by 1.18$\times$ on average (up to 1.72$\times$).  
In QAOA, node count directly maps to the number of qubits, while edges determine the number of entangling operations in the ansatz, affecting both circuit depth and compilation cost.

As shown in Figure~\ref{fig:impact_graph}(c), Enigma increases average degree by 1.1x on average (up to 1.58x).  
Average degree is defined as $\bar{d}=2E/N$, where $E$ and $N$ denote the number of edges and nodes.  
Density, defined as $D=2E/(N(N-1))$, increases by 1.02x on average (up to 1.45x), as shown in Figure~\ref{fig:impact_graph}(d).  
Average degree captures the mean connectivity of each variable, while density measures global connectivity relative to the graph’s maximum.  
In QAOA, higher average degree implies more entangling gates per qubit, which increases both compilation overhead and error accumulation.  
Similarly, higher density corresponds to denser interaction patterns that often translate into deeper circuits or more complex embeddings, whereas sparsification can ease hardware mapping.

\subsubsection{Impact on QEC}
Enigma affects $T$ gates in two distinct ways.   
First, we observe that coefficient manipulation (e.g., \texttt{ValueGuard}) alters $R_z$ rotation angles in the QAOA ansatz but does not change the resulting $T$-gate counts.  
This observation is consistent with prior work~\cite{hao2025reducing} showing that SOTA Clifford+$T$ synthesis methods such as Gridsynth~\cite{ross2014optimal} depend only on the target precision, not the specific angle.  

\begin{figure}[h]
    \centering        
	\includegraphics[width=0.6\columnwidth]{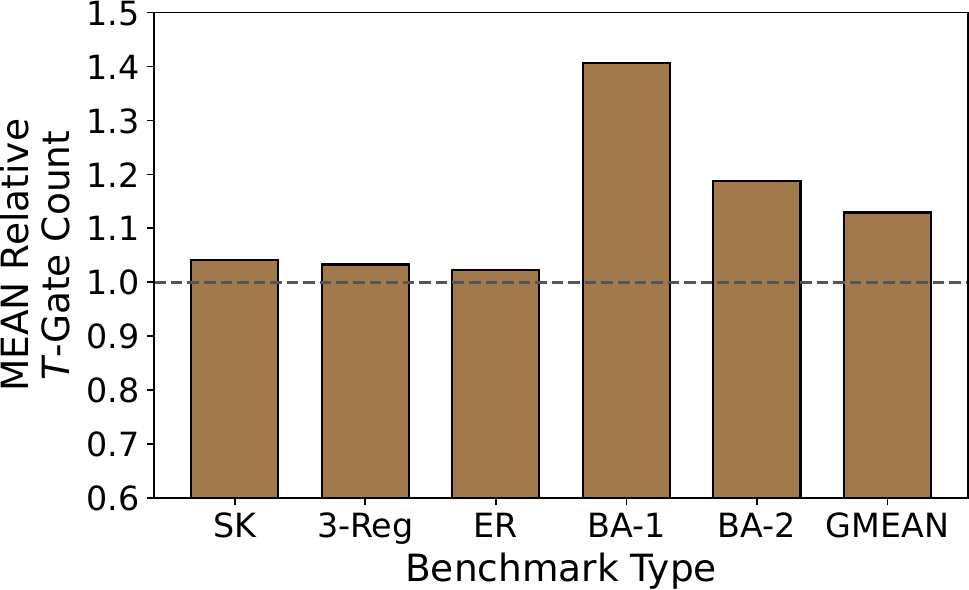}	
    \caption{
        Relative $T$-count under Enigma compared to the unobfuscated baseline (lower is better).
    }    
    \label{fig:impact_t_gate}
\end{figure}

Second, structural obfuscation (e.g., \texttt{StructureCamouflage}, \texttt{Regularizer}) introduces additional nodes and edges, which map to new operations and directly increase $T$-gate counts.  
As shown in Figure~\ref{fig:impact_t_gate}, Enigma increases $T$-count by 1.12$\times$ on average (up to 1.4$\times$).  
This increase remains modest relative to the protection gained, keeping QEC overhead manageable.

\subsubsection{Impact on Program Fidelity}
\begin{figure}[h]
    \centering        
	\includegraphics[width=\columnwidth]{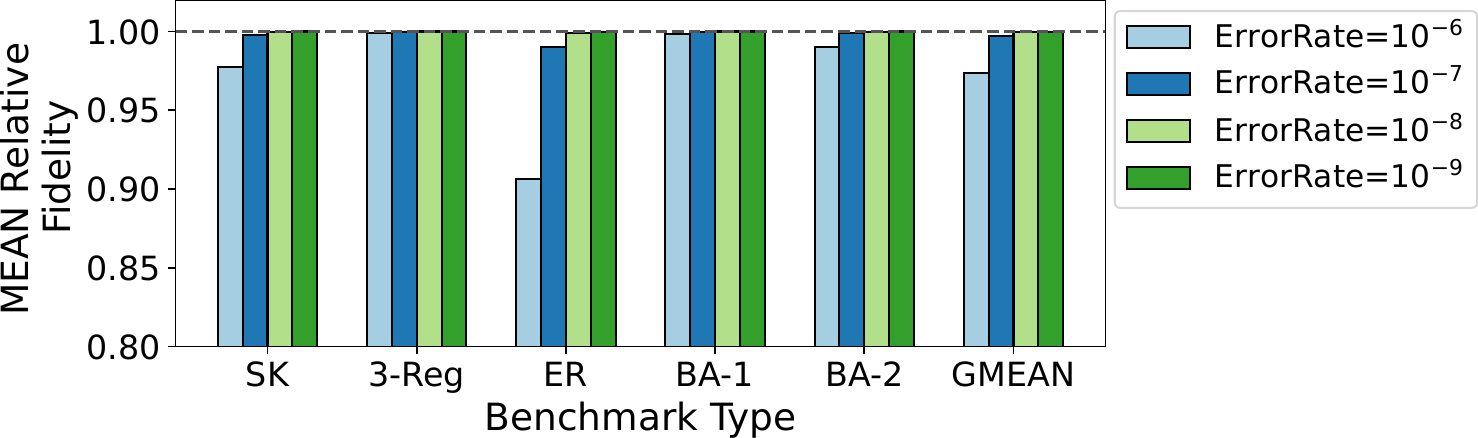}	
    \caption{        
        Relative program fidelity of Enigma compared to baseline across error rates $10^{-6}$–$10^{-9}$ (higher is better).
    }    
    \label{fig:impact_pst}
\end{figure}

We evaluate program fidelity by applying realistic error rates ranging from $10^{-6}$ to $10^{-9}$, reflecting expectations in the EFT era.  
As shown in Figure~\ref{fig:impact_pst}, fidelity of obfuscated programs consistently approaches that of the baseline as error rates decrease.  
Even at a conservative error rate of $10^{-6}$, the average relative fidelity exceeds 0.97, and at $10^{-7}$ or lower the gap is nearly negligible.  
These results indicate that Enigma introduces only modest fidelity overhead, which becomes practically insignificant under projected EFT error rates.

\subsubsection{Runtime Overhead}  
Table~\ref{tbl:obfuscation_time} reports Enigma’s runtime across graph types.  
The average obfuscation time is 5.44~s with variation by topology.  
Results are obtained using a single-threaded Python implementation and represent an upper bound.  
Reconstruction, in contrast, depends mainly on problem size rather than topology and consistently completes in under one second for all graph types.

\begin{table}[h]
\centering
\caption{
    Avg. obfuscation time (sec) across graph types.
    }
\begin{tabular}{c c c c c c}
\toprule
SK & 3-Reg & ER & BA-1 & BA-2 & Mean \\
\midrule
18.38 & 0.27 & 7.74 & 0.39 & 0.41 & 5.44 \\
\bottomrule
\end{tabular}
\label{tbl:obfuscation_time}
\end{table}

\section{Related Work}

Secure Quantum Computing addresses the delegation of quantum programs to untrusted servers, with approaches broadly categorized into Blind Quantum Computing (BQC) and Quantum Homomorphic Encryption (QHE).  
Most BQC protocols~\cite{broadbent2009universal,morimae2013blind,li2021blind,fitzsimons2017private,fitzsimons2017unconditionally,childs2001secure} require quantum capabilities on the client side, relying on quantum networks that are not yet widely available.  
Multi-server BQC variants~\cite{reichardt2013classical,huang2017experimental,mckague2013interactive,gheorghiu2017rigidity,gheorghiu2015robustness,fitzsimons2017private} remove this requirement but depend on non-communicating “untrusted” servers.  
QHE~\cite{broadbent2015quantum,fisher2014quantum,broadbent2015delegating,dulek2016quantum,ouyang2018quantum,tan2016quantum,yu2014limitations,liang2013symmetric} works with a single server but incurs exponential overhead and requires advanced quantum error correction.  

Enigma avoids these limitations by introducing adaptive, graph-aware obfuscation with polynomial overhead, making it practical for current and near-term systems.  
It is agnostic to algorithm (e.g., QAOA or future optimizers), computing model (e.g., gate-based, adiabatic, and one-way ), and qubit technology, and can run on quantum accelerators such as quantum annealers.  
Unlike recent secure hardware designs for superconducting systems~\cite{trochatos2023hardware,trochatos2023quantum}, which may not protect structured circuits, Enigma protects quantum optimization workloads across heterogeneous quantum platforms.

\section{Conclusion}

This work introduces \emph{application-specific secure quantum computing}, which brings obfuscation to the application layer to enable practical protection of sensitive workloads in the Early Fault-Tolerant era.  
We introduce \emph{Enigma}, the first realization of this principle for quantum optimization, that protects details of quantum optimization problem, including coefficients, variables, and graph structure while ensuring solution recovery.  
Evaluated against powerful AI-driven adversaries across diverse graph families, Enigma demonstrates strong resilience, with adversaries succeeding in only 4.4\% of cases.  
These protections incur modest overhead, with problem size and $T$-count increasing by just 1.07$\times$ and 1.13$\times$ on average, while obfuscation and decoding complete within seconds even for large-scale inputs.  

\textbf{
By aligning security directly with workload structure, this work establishes a new foundation for protecting quantum optimization in the EFT era.  
Looking ahead, application-specific SQC offers a scalable and adaptable pathway toward secure quantum computing across evolving algorithms, models, and hardware platforms, ensuring that quantum utility can be achieved without compromising trust.  
}


\newpage
\section*{Appendix}
\appendix

\section{Equivalence and Conversion Between Ising and QUBO Formulations}
\label{apx:ising_qubo}

\begin{figure*}[t]    
    \centering        
    \includegraphics[width=0.9\textwidth]{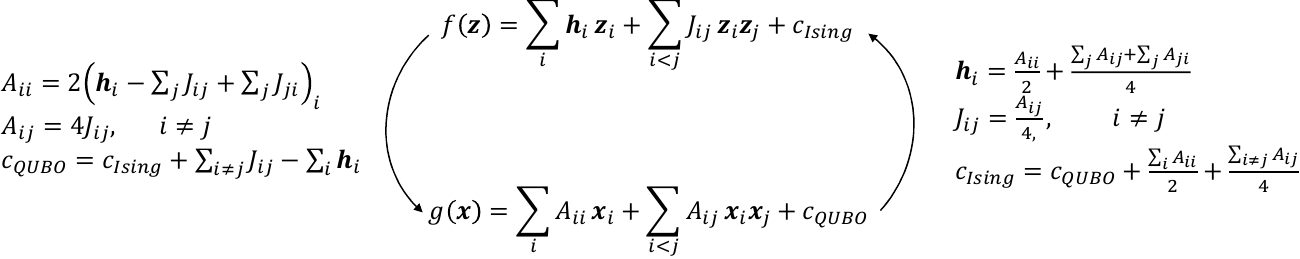}
    \caption{
        Linear transformability between the Ising Hamiltonian, $f(\mathbf{z})$ with $\mathbf{z}_i \in \{-1,+1\}$, and the Quadratic Unconstrained Binary Optimization (QUBO) formulation, $g(\mathbf{x})$ with $\mathbf{x}_i \in \{0,1\}$.  
        The diagram illustrates the linear conversion process in both directions between these two equivalent representations.  
    }
    \label{fig:ising_qubo}
\end{figure*}

The \emph{Quadratic Unconstrained Binary Optimization} (QUBO) problem is an optimization formulation over binary variables $\mathbf{x} = (\mathbf{x}_1, \mathbf{x}_2, \dots, \mathbf{x}_n)$, where $\mathbf{x}_i \in \{0,1\}$.  
Its general form is  
\begin{equation}
    g(\mathbf{x}) = \mathbf{x}^\mathsf{T} A \mathbf{x} + \mathbf{b}^\mathsf{T} \mathbf{x} + c_{\text{QUBO}},
\end{equation}
where $A \in \mathbb{R}^{n \times n}$ is an upper-triangular coefficient matrix, $\mathbf{b} \in \mathbb{R}^n$ contains the linear coefficients, and $c_{\text{QUBO}} \in \mathbb{R}$ is a constant offset term.  

The \emph{Ising} model is formulated over spin variables $\mathbf{z} = (\mathbf{z}_1, \mathbf{z}_2, \dots, \mathbf{z}_n)$, where $\mathbf{z}_i \in \{-1, +1\}$.  
Its objective function is  
\begin{equation}
    f(\mathbf{z}) = \sum_{i=1}^n \mathbf{h}_i \mathbf{z}_i + \sum_{i<j} J_{ij} \mathbf{z}_i \mathbf{z}_j + c_{\text{Ising}},
\end{equation}
where $\mathbf{h} \in \mathbb{R}^n$ contains the linear coefficients, $J \in \mathbb{R}^{n \times n}$ contains the quadratic interaction terms, and $c_{\text{Ising}} \in \mathbb{R}$ is a constant offset.  

The two formulations are \emph{mathematically equivalent} ~\cite{boettcher2019analysis,lucas2014ising,ayanzadeh2020ensemble,ayanzadeh2024skipper,ayanzadeh2020reinforcement,ayanzadeh2022quantum,ayanzadeh2020leveraging} under the linear change of variables  
\begin{equation}
    \mathbf{z}_i = 2\mathbf{x}_i - 1, \quad \text{for} \quad i = 1, \dots, n.
\end{equation}
Substituting this transformation into one model yields the other, establishing a direct mapping between $(\mathbf{h}, J, c_{\text{Ising}})$ in the Ising form and $(A, \mathbf{b}, c_{\text{QUBO}})$ in the QUBO form.  

This equivalence is important because QUBO can be expressed compactly in matrix form, making it well-suited for linear-algebra-based techniques.  
Such a representation is convenient for scaling, permutation, and structured perturbations while preserving problem structure.  

Figure~\ref{fig:ising_qubo} illustrates the bidirectional mapping between Ising and QUBO models.  
Because of this equivalence, all obfuscation techniques proposed in this paper can be applied regardless of whether the problem is provided in Ising or QUBO form.

\begin{tcolorbox}[colback=blue!12]
The equivalence between the Ising and QUBO formulations ensures that all obfuscation techniques proposed in this paper can be applied regardless of whether the input is provided in Ising or QUBO form.
\end{tcolorbox}

\section{Decoy Nodes and Edges in QUBO Form}
\label{apx:decoy_var}

\subsection{Adding Decoy Nodes and Edges} 
\label{apx:appending_decoy_nodes}

To perturb the problem structure, we append $m$ \emph{decoy variables} $\mathbf{y} \in \{0,1\}^m$ to the original $n$ binary variables $\mathbf{x} \in \{0,1\}^n$, producing an $(n+m)$-dimensional vector
\begin{equation}
    \tilde{\mathbf{x}} \;=\;
    \begin{bmatrix}
        \mathbf{x}_1, \dots, \mathbf{x}_n, \mathbf{y}_1, \dots, \mathbf{y}_m
    \end{bmatrix}^\mathsf{T}.
    \label{eqn:x_tilde}
\end{equation}

Simply adding decoy variables is insufficient; to make them indistinguishable from primary variables, we introduce \emph{decoy edges} both between primary and decoy variables and among decoy variables themselves.  
Let $A \in \mathbb{R}^{n \times n}$ be the QUBO matrix of the original problem
\begin{equation}
    g(\mathbf{x}) \;=\; \mathbf{x}^\mathsf{T} A \mathbf{x} + c_{\text{QUBO}}.
    \label{eqn:x_A_x}
\end{equation}
We form the augmented coefficient matrix
\begin{equation}
    \tilde{A} \;=\;
    \begin{pmatrix}
        A & B \\
        B^\mathsf{T} & C
    \end{pmatrix},
    \label{eqn:A_tilde}
\end{equation}
where  
$B \in \mathbb{R}^{n \times m}$ encodes interactions between $\mathbf{x}$ and $\mathbf{y}$, and  
$C \in \mathbb{R}^{m \times m}$ encodes interactions among $\mathbf{y}$ variables.  

The obfuscated objective becomes
\begin{equation}
    \tilde{g}(\tilde{\mathbf{x}}) \;=\; \tilde{\mathbf{x}}^\mathsf{T} \tilde{A} \tilde{\mathbf{x}} + c_{\text{QUBO}},
    \label{eqn:xt_At_xt}
\end{equation}
which can be expanded as
\begin{equation}
    \tilde{g}(\mathbf{x}, \mathbf{y}) \;=\; \mathbf{x}^\mathsf{T} A \mathbf{x} \;+\; 2\,\mathbf{x}^\mathsf{T} B \mathbf{y} \;+\; \mathbf{y}^\mathsf{T} C \mathbf{y} \;+\; c_{\text{QUBO}}.
    \label{eqn:expanded}
\end{equation}

\subsection{Guaranteeing Recovery of the Original Optimum} 
\label{apx:successful_recovery}

Let $\mathbf{x}^*$ be any global minimizer of the original QUBO problem~\eqref{eqn:x_A_x}, and let $(\mathbf{x}', \mathbf{y}')$ be a global minimizer of the augmented problem~\eqref{eqn:xt_At_xt}.  
Our goal is to choose $B$ and $C$ so that $\mathbf{x}' = \mathbf{x}^*$ for all such minimizers, regardless of $\mathbf{y}'$.

Because the decoy variables $\mathbf{y}$ are fictitious, their values do not affect the true solution; we only require that their inclusion does not alter the optimal $\mathbf{x}$.  
For fixed $\mathbf{x}$, define
\begin{equation}
    F_{\mathbf{x}}(\mathbf{y}) \;=\; 2\,\mathbf{x}^\mathsf{T} B \mathbf{y} \;+\; \mathbf{y}^\mathsf{T} C \mathbf{y}.
    \label{eqn:Fx}
\end{equation}

\begin{lemma}[Non\mbox{-}negative decoy weights guarantee recovery]
\label{lem:nonneg_guarantee}
Suppose $B \ge 0$ elementwise and $C \ge 0$ elementwise.  
Then for every $\mathbf{x} \in \{0,1\}^n$, the minimizer of $F_{\mathbf{x}}(\mathbf{y})$ over $\mathbf{y} \in \{0,1\}^m$ is attained at $\mathbf{y} = \mathbf{0}$.  
Consequently, any global minimizer of $\tilde{g}$ has the form
\begin{equation}
    \tilde{\mathbf{x}}^* \;=\; 
    \begin{bmatrix}
        \mathbf{x}^* \\[2pt]
        \mathbf{0}
    \end{bmatrix}
\end{equation}
where $\mathbf{x}^*$ is a global minimizer of $g$.
\end{lemma}

\begin{proof}
Fix $\mathbf{x} \in \{0,1\}^n$.  
If $B \ge 0$ elementwise, then for any $\mathbf{y} \neq \mathbf{0}$, we have $\mathbf{x}^\mathsf{T} B \mathbf{y} \ge 0$.  
If $C \ge 0$ elementwise, then $\mathbf{y}^\mathsf{T} C \mathbf{y} \ge 0$ for all $\mathbf{y}$.  
Thus $F_{\mathbf{x}}(\mathbf{y}) \ge 0$ for all $\mathbf{y}$, with equality only at $\mathbf{y} = \mathbf{0}$.  
Therefore $\mathbf{y} = \mathbf{0}$ is always a minimizer of~\eqref{eqn:Fx}.  

Substituting $\mathbf{y} = \mathbf{0}$ into~\eqref{eqn:expanded} yields
\begin{equation}
    \min_{\mathbf{y} \in \{0,1\}^m} \tilde{g}(\mathbf{x}, \mathbf{y}) \;=\; g(\mathbf{x}).
\end{equation}
Minimizing over $\mathbf{x}$ recovers exactly the original optimum $\mathbf{x}^*$.  
\end{proof}

This lemma implies that by sampling decoy edge weights from a non\mbox{-}negative distribution for both $B$ and $C$, we guarantee that the optimal primary variables are preserved exactly after augmentation.

\begin{tcolorbox}[colback=blue!12]
A sufficient condition in the QUBO (binary) domain for guaranteeing that the original solution can be recovered from the obfuscated problem is that all added linear and quadratic coefficients involving decoy variables are non-negative.
\end{tcolorbox}

\section{Trimming Spin Variables}
\label{apx:FQ}

\subsection{Fixing Values}
Suppose we \emph{trim by fixing} variable $z_k$ to either $+1$ or $-1$. This yields two reduced sub-problems:
\[
f^{z_k=+1}(z_{\setminus k}) = f(z_1,\ldots,z_{k-1},+1,z_{k+1},\ldots,z_N),
\]
\[
f^{z_k=-1}(z_{\setminus k}) = f(z_1,\ldots,z_{k-1},-1,z_{k+1},\ldots,z_N),
\]
where $z_{\setminus k}$ denotes the vector of all variables except $z_k$. Each sub-problem has $N-1$ variables.

\noindent{\bf{Recursive Freezing}}
By freezing $m$ variables, we obtain
\[
2^m \quad \text{sub-problems, each with } N-m \text{ variables.}
\]

\begin{figure}[t]
    \centering        \includegraphics[width=0.7\columnwidth]{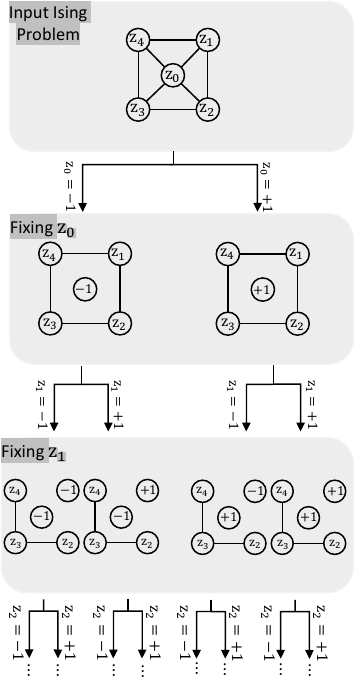}	
    \caption{
        Trimming Spin Variables. Nodes $Z_0$ and $Z_1$ are fixed, resulting in 4 smaller graphs.
    }    
    \label{fig:FQ_overview}
\end{figure}

\noindent{\bf{Symmetry Reduction}}
If the objective function satisfies the natural symmetry
\[
f(z) = f(-z),
\]
which is the case of Ising models with zero linear coefficients, then solutions come in symmetric pairs. That is, if $z^*$ is a global optimum, then so is $-z^*$. This allows us to evaluate only half of the sub-problems, inferring the results for their symmetric counterparts. Figure~\ref{fig:FQ_overview} shows an example of fixing two variables.

\noindent{\bf{Algorithm Outline}}
The trimming procedure can be summarized as:
\begin{enumerate}
    \item Identify variables (e.g., those with the highest degree in the problem graph).
    \item Fix one or more of these variables, generating smaller sub-problems $f^{z_k=\pm 1}$.
    \item Compile a single template quantum circuit and reuse it across all sub-problems by adjusting parameters.
    \item Run the quantum optimization algorithm (e.g., QAOA) on each reduced instance.
    \item Exploit symmetry to avoid redundant computations.
    \item Select the best solution among the sub-problems as the final output.
\end{enumerate}

\subsection{Guaranteed Recovery}
\subsection*{Guaranteed Recovery}

An important property of the fixing-based decomposition is that it ensures
\emph{guaranteed recovery} of the global optimum of the original problem.
Formally, let
\[
z^* = \arg\min_{z \in \{-1,+1\}^N} f(z)
\]
denote the global minimizer of $f(z)$. Suppose we fix a variable $z_k$ and
consider the two reduced subproblems
\[
f^{z_k=+1}(z_{\setminus k}), \quad f^{z_k=-1}(z_{\setminus k}).
\]
By construction, the global optimum $z^*$ must lie in exactly one of these two
cases, since $z_k^* \in \{-1,+1\}$. Thus,
\[
\min_{z \in \{-1,+1\}^N} f(z)
= \min\Big( \min_{z_{\setminus k}} f^{z_k=+1}(z_{\setminus k}), \;
             \min_{z_{\setminus k}} f^{z_k=-1}(z_{\setminus k}) \Big).
\]

Extending this argument recursively, if $m$ variables are fixed, then the
global optimum must appear in one of the $2^m$ subproblems. Therefore,
evaluating all subproblems (or their symmetric representatives) guarantees that
the true global optimum of the original function $f(z)$ is recovered.

\bibliographystyle{ACM-Reference-Format}
\bibliography{refs}

\end{document}